# Colloidal Two-dimensional Metal Chalcogenides: Realization and Application of the Structural Anisotropy

Ziyi Hu[1,2,†], Ryan O'Neill[1], Rostyslav Lesyuk[3,4], Christian Klinke[1,3,5,*]

[1] Chemistry Department, Swansea University - Singleton Park, Swansea SA2 8PP, United Kingdom
[2] Department of Physics, University of Warwick, Coventry CV4 7AL, United Kingdom
[3] Institute of Physics, University of Rostock, Albert-Einstein-Strasse 23, 18059 Rostock, German
[4] Pidstryhach Institute for Applied Problems of Mechanics and Mathematics, National Academy of Sciences of Ukraine, 79060 Lviv, Ukraine; Lviv Polytechnic National University, Department of Photonics, 79000 Lviv, Ukraine
[5] Department "Life, Light & Matter", University of Rostock, Albert-Einstein-Strasse 25, 18059 Rostock, Germany

* Corresponding author: christian.klinke@uni-rostock.de

**CONSPECTUS:** Due to the spatial confinement, two-dimensional metal chalcogenides display an extraordinary optical response and carrier transport ability. Solution-based synthesis techniques such as colloidal hot injection and ion exchange provide a cost-effective way to fabricate such low-dimensional semiconducting nanocrystals. Over the years, developments in colloidal chemistry made it possible to synthesize various kinds of ultrathin colloidal nanoplatelets, including wurtzite- and zinc blende-type CdSe, rocksalt PbS, black phosphorus-like SnX (X=S or Se), hexagonal copper sulfides, selenides and even transition metal dichalcogenides (TMD) like $MoS_2$. By altering experimental conditions and applying capping ligands with specific functional groups, it is possible to accurately tune the dimensionality, geometry and consequently the optical properties of these colloidal metal chalcogenide crystals. Here, we review recent progresses in the syntheses of two-dimensional colloidal metal chalcogenides (CMCs) and property characterizations based on optical spectroscopy or device-related measurements. The discoveries shine a light on their huge prospect for applications in areas such as photovoltaics, optoelectronics and spintronics. In specific, the formation mechanisms of two-dimensional CMCs are discussed. The growth of colloidal nanocrystals into a two-dimensional shape is found to require either an intrinsic structural asymmetry or the assist of coexisted ligand molecules, which act as lamellar double-layer templates or 'facet' the crystals via selective adsorption. By performing optical characterizations and especially ultrafast spectroscopic measurements on these two-dimensional CMCs, their unique electronic and excitonic features are revealed. A strong dependence of optical transition energies linked to both inter-band and inter-subband processes on the crystal geometry can be verified, highlighting a tremendous confinement effect in such nanocrystals. With the self-assembly of two-dimensional nanocrystals or coupling of different phases by growing heterostructures, unconventional optical performances such as charge transfer state generation or efficient Förster resonance energy transfer (FRET) are discovered. The growth of large-scale individualized PbS and SnS nanosheets can be realized by facile hot injection techniques, which gives the opportunity to investigate the charge carrier behavior within a single nanocrystal. Based on results of the device-based measurements on these individualized crystals, structure asymmetry-induced anisotropic electrical response and Rashba effect caused by a splitting of spin-resolved bands in the momentum space due to strong spin-orbit-coupling (SOC) are demonstrated. It is foreseen that such geometry-controlled, large-scale two-dimensional CMCs can be ideal materials used for designing high-efficiency photonics and electronics.

KEY REFERENCES

- Schliehe, C.; Juarez, B. H.; Pelletier, M.; Jander, S.; Greshnykh, D.; Nagel, M.; Meyer, A.; Foerster, S.; Kornowski, A.; Klinke, C.; Weller, H. Ultrathin PbS Sheets by Two-Dimensional Oriented Attachment. *Science* **2010**, *329*, 550–553.[1] *Two-dimensional oriented attachment of lead sulfide (PbS) nanocrystals into ultrathin single-crystalline micron-scale nanosheets is achieved by colloidal hot-injection synthesis.*

- Gerdes, F.; Navio, C.; Juarez, B. H.; Klinke, C. Size, Shape, and Phase Control in Ultrathin CdSe Nanosheets. *Nano Lett.* **2017**, *17*, 4165–4171.[2] *Syntheses of ultrathin CdSe nanosheets with tunable sizes (from tens of nm to over 100 nm), shapes (from sexangular to quadrangular to triangular) and phases (from zinc blende (ZB) to wurtzite (WZ)) are reported.*

- Dai, L.; Lesyuk, R.; Karpulevich, A.; Torche, A.; Bester, G.; Klinke, C. From Wurtzite Nanoplatelets to Zinc Blende Nanorods: Simultaneous Control of Shape and Phase in Ultrathin ZnS Nanocrystals. *J Phys. Chem. Lett.* **2019**, *10*, 3828–3835.[3] *Ultrathin ZnS nanoplatelets (NPLs) with genuine excitonic absorption and emission performances are shown and their electronic fine structure is resolved based on the ab-initio calculations.*

- Moayed, M. M. R.; Bielewicz, T.; Zollner, M. S.; Herrmann, C.; Klinke, C. Towards Colloidal Spintronics through Rashba Spin-Orbit Interaction in Lead Sulphide Nanosheets. *Nat. Commun.* **2017**, *8*, 15721.[4] *Rashba spin–orbit interaction in colloidal PbS nanosheets is demonstrated. The symmetry is broken due to quantum confinement, presence of asymmetric vertical interfaces and a gate electric field, which leads to the Rashba-type band splitting in momentum space and results in an unconventional selection mechanism for the excitation of carriers.*

1 INTRODUCTION

Two-dimensional metal chalcogenides are regarded as a class of promising semiconductors with promising electronic performances.[5,6] The large quantum confinement along their thickness direction as well as a large dielectric contrast give rise to tremendous exciton binding energies (for colloidal CdSe nanoplatelets, it goes up to several hundreds of meV) and thus contribute to unique optical and electronic performances.[7] Traditional fabrication techniques like vapor deposition, molecular beam epitaxy and mechanical exfoliation usually suffer from a poor scalability and have difficulties in supplying a large quantity of nanomaterials within short periods. In the recent years, the enormous development in colloidal chemistry approaches helped to overcome this disadvantage. By growing nanocrystals in a liquid phase, efficient control on materials' size and shape by changing synthesis conditions (e.g., reaction temperature, precursor concentration) becomes possible. The liquid phase-grown nanocrystals, which are known as 'colloidal nanocrystals', usually display a controllable size and shape. The growth of two-dimensional metal chalcogenides in solution can be realized by either heating-up (e.g., hydrothermal or solvothermal synthesis)[8,9] or a hot injection method.[10] According to the number of publications in the last 25 years, hot injection has been developed as a standard approach for the synthesis of CMCs. Compared to the heating-up method, where monomers are produced in a continuous fashion, hot injection enables monomer generation within a short period, leading to a separation of nucleation and particle growth phases. In such a way, inhomogeneous crystal growth can be effectively suppressed. Typical two-dimensional colloidal metal chalcogenides such as Cd-, Zn-, Pb-, Sn- and Cu-based chalcogenide nanoplatelets/nanodisks/nanosheets are now able to be grown by a one-pot hot injection process, which hugely reduces the costs. In this work, we review the recent progress in this field regarding both material synthesis and characterizations on their performances.

2 MECHANISMS OF THE ANISOTROPIC GROWTH

2.1 Generation of Physical Constraints

Lattice symmetry plays a significant role in the generation of two-dimensional geometries. Materials with an anisotropic crystal structure (herzenbergite-, covellite-, clockmannite- or paraguanajuatite-type germanium, tin, copper and bismuth chalcogenides etc.) possess intrinsic two-dimensional constraints, whilst II-VI (CdX, ZnX, HgX, X=S, Se, Te) and IV-VI (PbX) compounds with either a cubic or hexagonal lattice display a higher degree of symmetry. For crystals with an intrinsic asymmetric structure, the two-dimensional growth is achieved by the expansion of crystal along its predefined structural direction. An example for this is that the structurally asymmetric covellite CuS nanocrystals possess strong S-S bonds along the [0001] direction and thus favor Cu-S bond cleaving.[11] Consequently, the preferential growth orientation of such a crystal is perpendicular to its z-axis. Layered bismuth chalcogenides[12] possess a hexagonal crystal unit cell consisting of five atom layers, where weak bonds form between the layers due to van der Waals forces. This leads to an easy cleavage of the (0001) planes. For crystals with a relatively more symmetric lattice structure, such as face-centered cubic (fcc)-type CdSe or rocksalt-type PbS, the formation of a two-dimensional geometry is usually achieved by the "faceting" effect of ligand molecules.[1,13]

The formation pathways of colloidal two-dimensional nanocrystals can be described by one of the following mechanisms: (1) selective condensation of monomers on the target crystal facets; (2) orientational attachment of small crystals (e.g., magic-sized clusters); (3) lamellar ligand double layer-assisted growth and (4) ion exchange in 2D template.

The first formation pathway can often be observed in the synthesis process of two-dimensional noble metals such as silver, gold and palladium nanoplatelets/nanodisks.[14,15] Only a few types of metal chalcogenides which have an intrinsic anisotropic lattice structure are found to grow in such a manner. For most of the CMCs, the two-dimensional growth is achieved by either template-assisted crystal expansion or orientational attachment of crystals/clusters. Template-assisted synthesis is usually associated with the formation of self-organized double lamellar assemblies of long-chain aliphatic ligands (e.g., amines or fatty acids).

2.2 Synthesis of Two-Dimensional Colloidal Nanocrystals

CdSe nanocrystals are one of the most extensively studied CMCs in the last couples of decades. The bandgap of CdSe quantum dots varies from ~3.1 eV (398 nm) to 1.9 eV (656 nm) depending on their sizes (1.75-4.81 nm)[16] Dubertret's group pioneered the synthesis of zinc blende (ZB)-CdSe nanoplatelets. They managed to grow ultrathin CdSe nanocrystals by a simple hot injection experiment and achieved a precise control of the crystal geometrical parameters (i.e., thickness, lateral size, crystal shape).[15,17] In the following years, they adopted similar methods to synthesize colloidal CdS and CdTe nanoplatelets.[18] It was found that the position of excitonic absorption peak of these Cd-based nanoplatelets showed a heavy dependence on their thicknesses (or number of layers). A recent study, however, demonstrated that the lateral size (5–30 nm) of CdSe nanoplatelets can also affect their in-plane confinement and led to the exciton motion quantization.[19] In 2018, Ithurria et al. discovered that the CdSe nanoplatelets can be converted into shell-core-shell HgSe-CdSe-HgSe heteroplatelets by a simple cation exchange treatment without a change of their crystal structure (**Figure 1a**).[20] Apart from synthesizing heterostructures by cation exchange, Talapin and colleagues adopted a colloidal atomic layer deposition technique and converted the hot injection-grown CdSe nanoplatelets into shell-core-shell xCdS-CdSe-xCdS heteroplatelets (x is the number of shell layers and varies from 1 to 8) (**Figure 1b**).[21]

Wurtzite (WZ) CdSe nanocrystals were also obtained and studied by several other groups. Joo *et al.* synthesized WZ CdSe nanoribbons based on the reaction between $CdCl_2$ and octylammonium selenocarbamate in octylamine [22] (**Figure 1c**). The synthesized nanocrystals possessed a thickness of only 1.4 nm. Buhro's group discovered the formation of WZ CdSe nanoplatelets and two-dimensional CdSe belts in their solution-based synthesis experiments conducted at different temperatures.[23,24] Our group recently demonstrated that the use of halogenated alkanes gave rise to an efficient control of crystal geometry and phase composition of CdSe nanoplatelets[2] (**Figure 1d** and **e**). By adding different amount of 1-bromoheptane (1-Br-Hep) to the reaction solution, we observed a bi-phasic shape evolution (from sexangular to quadrangular and then to triangular) and a change of crystal phase from ZB to WZ. Based on the mass spectrometry and X-ray photo-electron spectroscopy (XPS), it was found that the 1-Br-Hep molecules and the free Br$^-$ ions released by them can replace the pristine carboxyl ligands (acetate) during the process of crystal growth. According to the density functional theory (DFT) calculations, the wurtzite phase was supposed to form when the nanocrystals were capped by acetate. This is in a good agreement with the observed phase evolution from ZB to WZ with the presence of halogenated alkanes.

The formation mechanism of CdSe nanoplatelets has been actively debated in the past decades. Riedinger *et al.* studied the growth of CdSe nanoplatelets in isotropic melts containing Cd(carboxylate)$_2$ and elemental Se.[25] According to both experimental observations and theoretical analyses (e.g., crystal growth modeling, kinetic Monte Carlo simulations), they found that the highly anisotropic two-dimensional shape arose from the intrinsic instability of the growth kinetics. It was also demonstrated that the nanoplatelet growth rate at the edges was much higher than it on the large lateral facets, favoring the growth of thicker nanocrystals at higher temperatures. The discoveries suggest that the formation of two-dimensional nanoplatelets in such melts was controlled by an island-nucleation-limited growth mechanism.

In 2017, Peng's group reported the formation of rectangular ZB-CdSe nanosheets with a thickness of 1.5 nm by hot injection synthesis.[26] It was demonstrated that the growth of two-dimensional crystals was triggered only when both short-length (acetate) and extended carboxylate ligands (stearate) were introduced to the synthesis solution. The formation of such nanosheets was assumed to be carried out in three phases: (1) Generation of single-dot intermediates with nearly flat {100} basal planes through crystal ripening. (2) Formation of two-dimensional embryos by oriented attachment of single-dot intermediates followed by inter-particle ripening. (3) Formation of rectangular CdSe nanosheets with six well-defined {100} facets based on the embryo-nanocrystal attachment followed by a third-step ripening.

Very recently, Castro *et al.* investigated the growth mechanism of colloidal ZB-CdSe nanoplatelets by synchrotron-based small-angle X-ray scattering (SAXS) and wide-angle X-ray scattering (WAXS).[27] According to their experimental results, various features were revealed from the time-resolved spectra, which were assigned to the generation of Cd(oleate)$_{2-x}$(acetate)$_x$ domains, crystallization of alkyl chains and lateral extension of nanoplatelets. It was thus verified that the CdSe nanoplatelets formed by the lateral expansion of small nuclei without the participation of soft ligand templates.

Warner *et al.* first reported the synthesis of triangular and hexagonal CdS nanoplates with a WZ phase in the alkylamine medium.[28] Cheng *et al.* optimized the synthesis parameters and obtained hexagonal WZ CdS nanoplatelets presenting a narrow lateral size distribution.[29] According to the results of transmission electron microscopic (TEM) imaging and selected area electron diffraction analysis, it was confirmed that the growth of the nanocrystal was carried out along the [100] direction (or lateral direction). Son *et al.* reported the synthesis of 1.2 nm-thick two-dimensional CdS nanoplates and found that the magic-size clusters were generated prior to the formation of two-dimensional crystals based on both TEM imaging and optical absorption spectroscopy.[30] It thus led to a conclusion that CdS nanoplates formed through the two-dimensional assembly of magic-size clusters. Stam *et al.* demonstrated that the colloidal Cu$_{2-x}$S nanoplatelets can be turned into CdS nanoplatelets by a cation exchange treatment.[31]

Among different types of metal chalcogenides, lead chalcogenides show a remarkable carrier mobility, which allows them to be a potential candidate for electronics and photonics such as IR detectors and solar cells. PbS nanocrystals with sizes below the bulk exciton Bohr radius limit $a_B$ (18 nm)[32] display both a strong electronic confinement and a decrease in the electron-phonon coupling strength.[33] Weller's group and us, we first reported the formation of PbS nanosheets based on a hot injection approach using 1,2-dichloroethane and oleic acid as co-ligands (**Figure 2**).[1,34] We verified that the micron-scale PbS nanosheets were synthesized by the spontaneous two-dimensional attachment of small PbS nanocrystals with highly reactive {110} facets. Meanwhile, the SAXS analysis revealed the existence of oleic acid bilayers between the stacked PbS nanosheets (**Figure 2c**). The growth mechanism of such PbS nanosheets was thus assumed to be the orientational attachment of zero-dimensional PbS nanocrystals assisted by the two-dimensional ligand templating.

Years later, Khan *et al.* synthesized elongated rectangular PbS nanosheets (1.8-2.8 nm) by a one-pot heating-up approach.[35] It was found that the change of reaction temperature can lead to an evolution of both the crystal shape and the lattice parameters. The PbS nanosheets with an orthorhombic structure were found to grow at 80°C, while nanowires and nanocubes can be generated at higher temperatures (90°C). Shkir *et al.* synthesized colloidal PbS nanosheets by a direct mix up of thiourea and the aqueous Pb precursor solution containing lead acetate, NaOH and sodium dodecyl sulfate.[36] All of the synthesized PbS nanosheets had a large lateral size (more than one micron) and a small thickness (less than 5 nm).

Beyond PbS nanosheets, several types of two-dimensional metal chalcogenide nanocrystals had also been demonstrated to be grown based on the mechanism of template-assisted orientational attachment. Lauth *et al.* reported the growth of colloidal InSe nanosheets inside the octadecylamine templates,[37] with the crystal thickness being restricted to about 5 nm. Stam *et al.* synthesized ultrathin Cu$_{2-x}$S (0≤x≤2) nanosheets [31] and verified that the nanocrystals formed within the ligand templates of 1-dodecanethiol based on SAXS characterization.[38] Recently, our group demonstrated the growth of wurtzite ZnS nanoplatelets (1.8 nm in thick) with the assist of ligand templates comprising both long-chain oleylamine and shorter-chain octylamine.[3,39]

Black phosphorus-like IV-VI compounds were demonstrated to have a striking anisotropic thermoelectric performance.[40] Orthorhombic SnX and GeX (X=S or Se) nanocrystals are two typical IV-VI compounds. These crystals display a zigzag atom arrangement in *b*-direction whilst an armchair arrangement in *c*-direction. In 2011, Vaughn *et al.* reported the growth of mesoscale single-crystalline SnSe nanosheets with a thickness of 8-30 nm using trioctylphosphine, hexamethyldisilazane and oleylamine as ligands.[41] Liu *et al.* then studied the influence of the ligands' molecular composition on the shape of SnSe[42] by testing different ligand combinations (Se-alkylphosphine and Se-alkyl). It was shown that a two-dimensional shape can be favored when phosphine–selenium precursors were used, which was attributed to a relatively low nucleation rate. Li et al. reported the growth of four-atom-thick SnSe nanosheets by a one-pot heating-up method.[43] Interestingly, the crystal synthesis in the absence of 1,10-phenanthroline led to a growth of SnSe nanocrystals with three-dimensional shapes. Recently, our group synthesized the untruncated and truncated square-like nanoplatelets[44] and nearly rectangular SnS nanosheets[45] based on the hot injection approach (**Figure 1 f** and **g**). It was found that all the side facets of the untruncated square-like nanoplatelet were in an isotropic {101} type. For the truncated square-like nanoplatelets, however, an extra pair of (100) side facets present. The mutual restriction effect of ligand molecules was confirmed by the DFT studies, which explained the formation of SnS nanosheets with these different shapes.

SnTe is a promising small-bandgap thermoelectric and topological material.[46] Bulk phase SnTe presents a rocksalt crystal structure (space group *Fm3m*). Recently, it was found that the one-unit cell-thick SnTe nanosheets fabricated by molecular beam epitaxy showed a decrease in the crystallographic symmetry, resulting in a lattice structure similar to SnSe.[47] *Li et al.* found that the optimized configuration of monolayered SnTe isolated from the bulk orthorhombic SnTe had a lower symmetry ($Pmn2_1$).[48] Recently, our group synthesized quasi-two-dimensional SnTe nanostripes with a lateral size of up to 6 microns and thicknesses below 30 nm using a one-pot hot injection method.[49] These SnTe nanostripes were found to display a cube-like rocksalt structure and show pronounced absorption features in the IR range. Despite the extensive studies on colloidal SnX nanosheets, finding ways to refine the crystal thickness and eliminate surface oxidation are still major tasks.

Recently, our group investigated the shape evolution of CuS nanocrystals under the kinetic-controlled growth condition.[50] We found that the triangular CuS nanoplatelets and nanosheets comprising Cu-rich (100) facets dominated in such a condition. Based on the DFT simulations, it was confirmed that the binding energies of amine and carboxyl groups with the Cu-rich (100) facets were higher than those with the S-rich facets. Depending on the reactivity of the used S precursor, the shape of nanocrystal can be tuned from triangles to hexagonal nanoprisms. This enabled an active control of the plasmon-driven absorption performance in the NIR region.

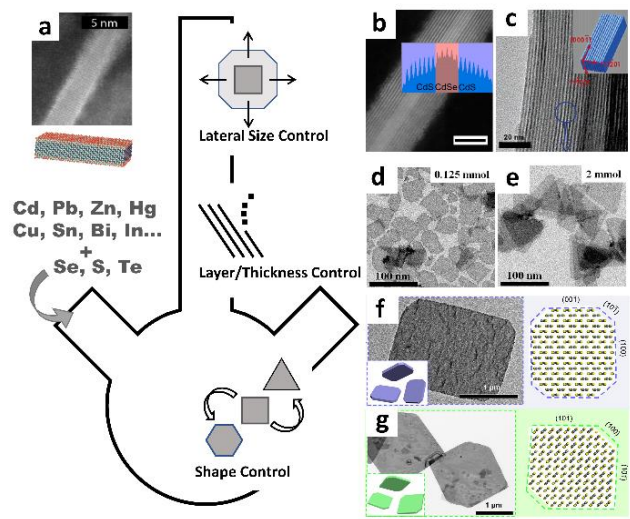

**Figure 1**: (a) High-angle angular dark field scanning transmission electron microscopic (HAADF-STEM) image and schematic of a 10 ML HgSe/CdSe/HgSe heterostructure. Reproduced with permission from ref.[20] Copyright 2018 American Chemical Society. (b) HAADF-STEM image of an 8ML CdS/CdSe/8ML CdS heterostructure nano-platelet (Inset: EDX line profile along the thickness direction). Reproduced with permission from ref.[21] Copyright 2014 American Chemical Society. (c) TEM image of stacked 1.4 nm-thick WZ-CdSe nano-ribbons. Reproduced with permission from ref.[22] Copyright 2006 American Chemical Society. (d, e) TEM images of CdSe nanoplatelets grown by the hot injection approach at different concentration of Br-ligand. The unit cell of orthorhombic SnX and GeX crystals. Reproduced with permission from ref.[2] Copyright 2017 American Chemical Society. (f, g) TEM images and schematics of (f) a near-rectangular and (g) truncated square-like SnS nanosheets. Reproduced with permission from ref.[45] Copyright 2019 American Chemical Society.

Besides the typical Cd-, Sn- and Cu-based chalcogenide crystals, hot injection method had also been applied to synthesize transition metal dichalcogenides (TMD). Jung *et al.* fabricated colloidal two-dimensional $MSe_2$ (M=W or Mo) nanocrystals by hot injection with the assist of oleylamine, oleyl alcohol or oleic acid.[51] The crystal growth could be tuned from the lateral orientation-preferred type (growth towards larger and single-layer crystals) to the vertical orientation-preferred type (growth towards smaller and multi-layer crystals) based on the choice of ligand molecules.

The growth mechanisms for some other types of two-dimensional CMCs are, however, still under debate. Berends *et al.* synthesized two-dimensional copper indium sulfide nanosheets based on hot injection experiments.[52] It was found that the existence of nonstoichiometry due to ligand molecules led to the anisotropic growth of as-synthesized small chalcopyrite nanocrystals. Nevertheless, in their study it was not shown whether the lateral expansion of crystals was driven by the templating effect of ligand bilayers or some other mechanisms. Almeida and co-workers realized the growth of monolayer-thick hexagonal *β*-$In_2Se_3$ nanosheets using the ionic precursor $lnCl_3$.[53] However, the mechanism of nanosheet formation has to be studied by future works.

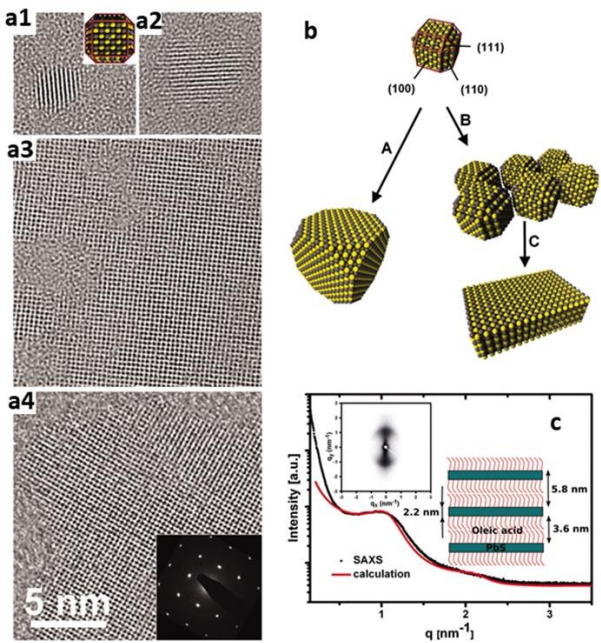

**Figure 2.** (a) TEM image of the colloidal PbS nanosheets at different growth stages. (a1) Ultrasmall PbS nanoparticle (truncated cuboctahedron). (a2, a3) Intermediate nanocrystal present during the process of two-dimensional attachment. (a4) Two-dimensional nanocrystal present at the final growth stage (insert is the electron diffraction pattern). (b) Schematic diagram showing the oriented attachment of tiny PbS nanoparticles into a large two-dimensional nanosheet. (c) Experimental and calculated SAXS of the OA-capped PbS nanosheets. Reproduced with permission from ref.[1] Copyright 2010 American Association for the Advancement of Science.

## 3 SPECTROSCOPIC CHARACTERIZATIONS

In two-dimensional CMC nanocrystals, the strong quantum confinement gives rise to a spatial overlap of the wavefunction of electrons and holes and an increase in the electron-hole Coulomb interaction.[54] This consequently alters the spectroscopic performances, which paves the way for the designing of high-efficiency optical and electronic devices.

### 3.1 Optical Response in a Broad Spectral Range

Because of the tunable effective band gaps, CMC nanocrystals show their optical responses over a broad spectral range from UV to NIR. Metal chalcogenides such as ZnS, CdSe and InSe display striking excitonic in the ultraviolet and visible wavelengths.

Relying on steady-state spectroscopy and theoretical modelling, the study of absorption features can yield precise information about crystal size, shape (or dimensionality) evolution during the synthesis. Traditionally for colloidal nanocrystals, effective-mass approximation had been employed to interpret the confinement effect and to quantify the exciton binding strength in CMC nanoplatelets[18].

The electronic structure of CdSe displays degenerate heavy hole (*hh*) and light hole (*lh*) valence bands, which have decreased effective masses.[55] Due to the selection rules, a heavy hole exciton forms an in-plane dipole and is therefore optically active (generating photon emission by recombination). Ithurria et al.[18] and Christodoulou et al.[56] investigated the band-edge emission of colloidal CdSe nanoplatelets with a thickness from 4 to up to 8.5 ML. It was confirmed that both the energy of electron-LH and electron-HH transition had a strong dependence on the crystal thickness (**Figure 3a**). Meanwhile, the fluorescent lifetime of these nanoplatelets showed a remarkable decrease with temperature, which was an indicator of giant oscillator strength transition (GOST). In addition to the excitonic features caused by the inter-band electron-HH and electron-LH transitions, Diroll et al. had recently demonstrated the absorption performance of colloidal CdSe nanoplatelets at near-infrared wavelengths due to inter-subband transition from $E_1$ to $E_2$ state.[57] Meanwhile, they found that the interaction of CdSe nanoplatelets with halides, thiolates, and phosphonates can induce bathochromic shifts of both the inter-band and inter-subband transitions.[58]

Due to the narrow bandgaps, Hg- and Pb-based CMCs display their optical features across the infrared window. The ultrathin PbS nanosheets synthesized by hot injection were found to display only a weak absorbance in the near-infrared region, which makes it hard to distinguish their excitonic features (**Figure 3b**).[35,59] Two-dimensional Hg-based CMCs, including HgSe and HgTe nanoplatelets as well as HgSe-coupled quantum well heterostructures, on the other hand, showed recognizable absorption features across the visible and near-infrared windows. It was demonstrated that the maximum of their absorbance related strongly to the crystal thickness.[20]

Bhandari et al.[60] and Khan et al.[35] characterized the fluorescent properties of ultrathin colloidal PbS nanosheets (**Figure 3b** and **3c**). It led to a conclusion that the emission wavelength of PbS nanosheets varied with the synthesis conditions (730-750 nm and 1400-2000 nm reported by Bhandari's and Khan's experiments, respectively). Jiang et al. investigated the fluorescence performance of the hot injection-grown colloidal PbS nanosheets at their different growth stages (small dots, intermediate product with a porous morphology and final nanosheets with smooth lateral facets).[61] A drastic shift of the fluorescent peak towards longer wavelengths was observed at the early stage of crystal growth, which can be interpreted by a decrease of the bandgap due to wave function extension upon the attachment of small dots into two-dimensional nanosheets. Also, the narrowing of fluorescent peak was observed when porous nanosheets evolved into nanosheets with smooth lateral surfaces.

Nanometer-scale binary copper-based chalcogenides (e.g., $Cu_{2-x}X$, X=S, Se or Te) show a 'self-doping' nature and can carry considerably more free carriers. This gives rise to the near-infrared localized surface plasmonic resonances (LSPRs). Consequently, an optical absorbance maximum locating at the LSPRs frequency of these nanocrystals is usually observed (**Figure 3d**).[62–64] The plasmonic properties of copper-based chalcogenides are widely discussed in several recent reviews.[65,66] Meanwhile, the optical nonlinearity of Cu-based chalcogenide nanocrystals had also been demonstrated, which sheds light on their potential application as an ultrafast pulse generator.[67]

Recently, we studied the one-dimensional confinement effect of ultrathin colloidal WZ ZnS nanoplatelets with the assist of DFT-based ab-initio calculations.[3] According to our modelling results, the transitions from hh-, lh- and split-off (so) valence bands were found in very close spectral proximity within one absorption peak due to a much smaller spin-orbit coupled effect in ZnS compared to CdSe (**Figure 3e**). In the experimental spectrum, an absorption peak ascribable to the transition from third valence band ($h_2$) can be resolved. The other absorption peaks presenting at higher energies were found related to the $h_4$–$e_1$, $h_1$-$e_3$ and $h_5$-$e_2$ transition, respectively.

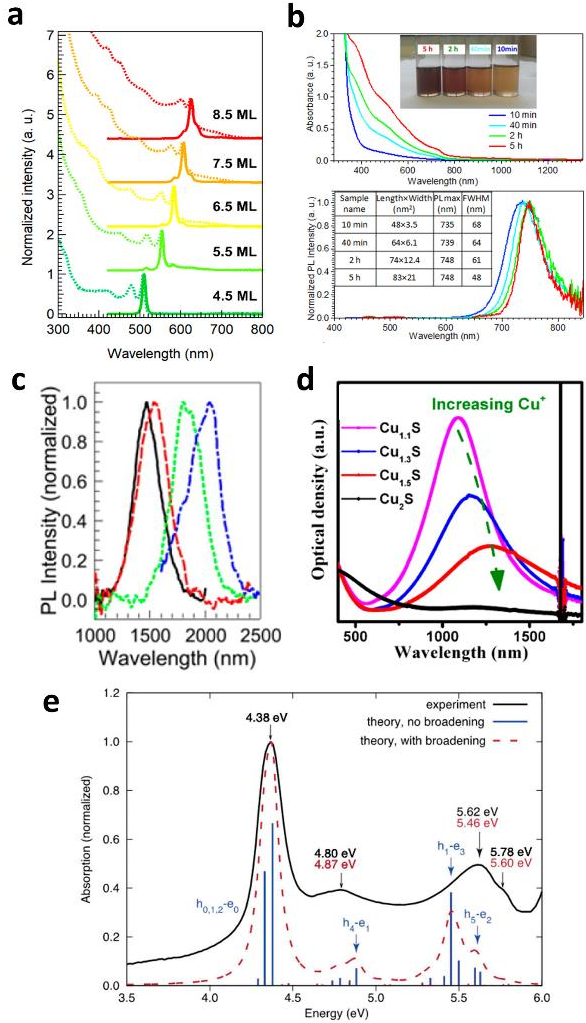

**Figure 3**: (a) The steady-state absorption and photoluminescence spectra of colloidal CdSe nanoplatelets with different thicknesses. Reproduced with permission from ref.[56] Copyright 2018 American Chemical Society. (b) The absorption and photoluminescence spectra of PbS nanosheets developed by Khan *et al.* for different growth durations. Reproduced with permission from ref.[35] Copyright 2017 American Chemical Society. (c) The absorption and photoluminescence spectra of PbS nanosheets developed by Bhandari *et al.* at different temperatures. Reproduced with permission from ref.[60] Copyright 2014 American Chemical Society. (d) Localized surface plasmon resonance and its damping in the optical extinction spectra of $Cu_xS$ nanocrystals. Upon addition of $Cu^+$ ions, stoichiometry varied from $Cu_{1.1}S$ (pink curve) to $Cu_2S$ (black curve). Reproduced with permission from ref.[63] Copyright 2013 American Chemical Society. (e) Absorption spectrum (black) of ZnS WZ nanoplatelets along with calculated dipole transitions (red dashed). Reproduced with permission from ref.[3] Copyright 2019 American Chemical Society.

## 3.2 Exciton Dynamics

Time-resolved spectroscopic characterizations offer a way to learn the exciton generation and recombination mechanisms within colloidal nanocrystals and in particular of two-dimensional CMCs. Talapin and colleagues studied the amplified spontaneous emission in colloidal CdSe nanoplatelets and quantified the contribution of Auger recombination to the relaxation of excited charge carriers based on time-resolved photoluminescence measurements (TRPL).[21] In another study, Demir and co-workers studied the spontaneous emission of CdSe nanoplatelets with different thicknesses.[68] By performing TRPL measurements on CdSe nanoplatelets under various temperatures (10–300 K), Achtstein et al. uncovered that the slow decay component increased with temperature.[54] This was found to be linked to the GOST effect as had been previously predicted for the two-dimensional system. Talapin's group also studied Förster resonance energy transfer (FRET) in a binary CdSe nanoplatelet system where 4 ML and 5 ML nanocrystals acted as the donor and the acceptor, respectively.[69] Based on their time-resolved spectroscopic measurements, it was revealed that FRET occurred at the picosecond timescale (6-23 ps). Such an energy transfer led to a more rapid photoluminescence decay of the donor at the presence of the acceptor whereas a slower decay of the acceptor at the presence of the donor. Guzelturk *et al.* investigated the photoluminescence decay and photoluminescence quantum yield of the self-assembled column-like CdSe nanoplatelet stacks, which both showed a drastic decrease with a rise in the degree of stacking.[70] In such a column structure, excitons can be transported between the close-packed nanoplatelets by FRET, which would be quenched when they encountered non-emissive nanoplatelets. For a deep understanding of the dynamics of photo-excited charge carriers in colloidal nanoplatelets, a combination of TRPL, transient absorption (TA), and terahertz (THz) spectroscopic techniques would be required.[71-73] Siebbeles et al. studied the decay dynamics of excitons in CdSe/CdS platelets and found that the band-to-band relaxation and the trap-induced decay occurred on a timescale of hundreds of picoseconds.[73] By means of optical pump–THz probe spectroscopy, they extracted the frequency-dependent alternating current mobility of colloidal Cd-based nanoplatelets, which displayed a significant decrease compared to the value of its bulk counterpart. This indicated that the quenching of excitons was induced by the hole trapping at quenching sites and continued movement of the electrons in the Coulomb potential of the trapped hole.

The assignment of dark and bright exciton energies in ZB-CdSe nanoplatelets was carried out by Shornikova *et al.* based on their TRPL measurements.[74] They showed that the charged excitons (trions) generation can give rise to extra emission lines. It was demonstrated that the low-energy emission line can be resolved at low temperatures, disregarding the thickness of nanocrystals (3, 4 and 5 ML).[75] Based on the TRPL measurements in the presence of an external magnetic field of up to 15 T, they determined the dynamics of trion radiative recombination (**Figure 4a** and **4b**). Relying on the TA measurements, Li and colleagues studied the exciton dynamics of type-II CdSe/CdTe core-crown nanoplatelets.[76] The measurement uncovered several photobleaching signals generated by both inter-band transitions ($T_1$-$T_4$) and a charge transfer state (CT) (**Figure 4c** and **4d**). Three spatially distinct excitons with increasing energies, the interface-localized exciton ($X_{CT}$), crown-localized exciton ($X_{CdTe}$), and core-localized exciton ($X_{CdSe}$), were determined from their experimental results. Since each single exciton level accommodated two excitons due to electron spin degeneracy, up to six exciton states could be generated (**Figure 4e–g**).

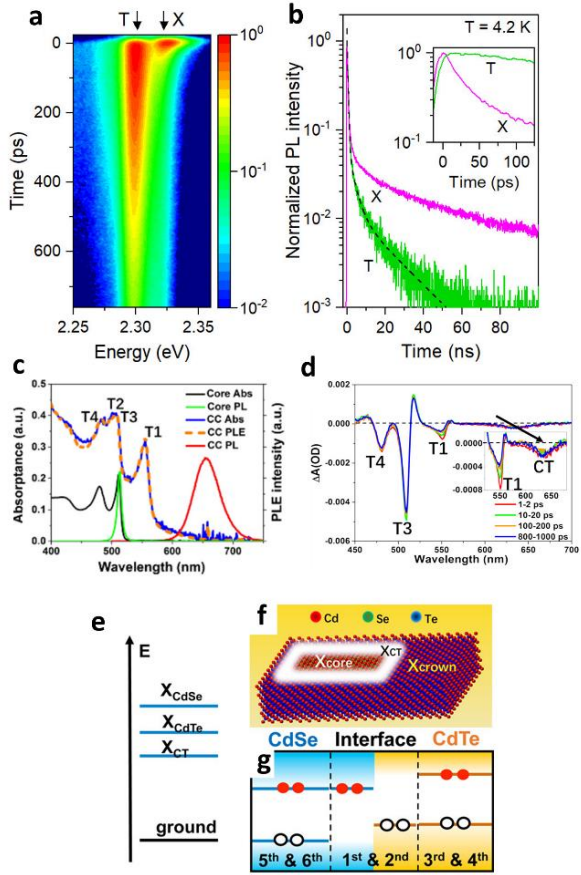

**Figure 4:** (a) Exciton and trion recombination dynamics of 5ML nanoplatelets revealed by time-resolved photoluminescence measurements at T = 4.2 K. (b) Photoluminescence decays of the exciton (pink) and trion (green) on different scales (ps and ns). Reproduced with permission from ref.[75] Copyright 2020 American Chemical Society. (c–g) Exciton dynamics of core-crown CdSe-CdS nanoplatelets revealed by transient absorption spectroscopy. Reproduced with permission from ref.[76] Copyright 2017 American Chemical Society. (c) The steady-state absorption spectra of CdSe core ("Core Abs") and core-crown CdSe-CdS ("CC Abs"), photoluminescence of CdSe core ("Core PL") and core-crown CdSe-CdS ("CC PL") and photoluminescence-excitation spectrum of core-crown CdSe-CdS ("CC PLE"). (d) The transient absorption spectra of core-crown CdSe-CdS taken from different time windows. The characteristics resulting from the excitonic transitions (T1, T3 and T4) and the charge transfer are identified. (e) Schematic diagram of the different exciton states in the core-crown CdSe-CdS nanoplatelets. (f,g) Three spatially separated exciton states in the core-crown nanoplatelets.

## 4. OPTOELECTRONIC PERFORMANCE OF INDIVIDUAL CMC NANOCRYSTALS

While showing a huge potential application of light emission, designing colloidal nanoplatelets and nanocrystal-based devices (e.g., transistors, solar cells) suffer from the existence of tunnel barriers and grain boundaries between the particles, which usually leads to a poor performance. Carrying out measurements on individual two-dimensional single-crystalline nanomaterials avoids such junctions between items and thus allows the characterization of materials' intrinsic properties. Electrical transport properties and photoresponsivities of individual colloidal two-dimensional nanocrystals are usually measured in transistor configuration. By changing parameters such as temperature $T$, source-drain voltage $V_{ds}$ and gate voltage $V_g$, it is possible to investigate various types of device performances including on/off ratio, capability of gate voltage-dependent carrier transport and carrier mobility.

In device-based measurements, the work function ($\varphi$) of the metal electrodes is key to realizing a low contact resistance and a high charge injection efficiency. A good Ohmic contact, in which case a low potential barrier at the metal-semiconductor interface (corresponding to a high charge injection rate) is formed, can be achieved if the work function of metal electrode is close to the highest occupied or lowest unoccupied state of semiconducting nanocrystals.[77] For instance, Au provides a large work function ($\varphi$ = 5.1 eV) that can be aligned with the $1S_h$ state of PbSe quantum dots (∼4.7 eV), leading to a low contact resistance and thus a higher electron and hole injection rate.[78] A high quantum efficiency is achieved by the rapid dissociation of photogenerated excitons and the transport of charges to electrodes without recombination. For colloidal nanocrystal-based devices, the charge carrier mobility can be enhanced by eliminating the trap / mid-gap states, altering doping levels or removing dangling bonds.

Ultrathin lead chalcogenide (PbS and PbSe) nanosheets performed efficient carrier multiplication for given photon energies, which yields additional electron-hole pairs as the photon energy exceeds a threshold. This enables a more efficient use of their band gap and makes them interesting for applications in optoelectronics and photovoltaics.[79] The electrical conductivity through ultrathin colloidal PbS nanosheets have been extensively studied by our group.[34,59,80] Based on field-effect transistor measurements, a remarkable mobility attributed to the strong confinement and effective filling of trap states was demonstrated.[34] Characterizations on individual PbS nanosheets with various thicknesses (4-20 nm) revealed the dependence of photovoltaic performance on crystals' geometrical parameters (**Figure 5a** and **5b**).[59] The individual ultrathin nanosheets yielded a power conversion efficiency of up to 0.94% when contacted with appropriate metal electrodes.[81] Recently, we investigated the effects of halogen ion doping and $O_2$ absorption on the electrical properties of PbS nanosheets.[80] With proper chemical treatment, the performance of colloidal nanosheets can even compete against that of exfoliated two-dimensional TMDs.

In the recent decades, the development of spintronic materials made it possible to employ and manipulate an extra degree of freedom provided by the spin of electrons. We demonstrated the Rashba effect, a splitting of the spin-resolved bands in k-space upon introduction of an asymmetry in the crystal potential and strong spin-orbit coupling (SOC),[82] in the two-dimensional PbS nanosheets.[4] This phenomenon is associated with the breaking of structural inversion symmetry break generated upon the formation of quantum wells, or the creation of asymmetry at nanocrystal edges and contacts. In our case, the symmetry was reduced by the two-dimensionality and the applied electrical gate field. Owing to the asymmetric distribution of carriers in momentum space in the presence of Rashba SOC, a spin-polarized net current can be detected upon exposure to circularly polarized laser light based on the measurement of the circular photo-galvanic effect (CPGE)[4] (**Figure 5c** and **5d**). The realization of such spin-related electrical transporting properties in colloidal nanomaterials opened a promising pathway towards the design of inexpensive spintronic devices.

Due to the anisotropic structure, black phosphorus-like SnX (X = S or Se) crystals exhibit a quantitative difference in their optical response and electrical conductivity in relation to their crystallographic growth directions. The anisotropic electrical transport and photoresponse of colloidal two-dimensional SnS nanosheets had been studied by various research groups.[45,83] Biacchi *et al.* characterized the inherent transport property of colloidal SnS nanoribbons and nanosheets based on a multi-probe setup, which enabled the characterization of conductivity along two directions set off 90° from each other.[83] For SnS nanoribbons, a lower contact resistance was obtained when probing along the ribbon length direction, which corresponded to the *b*-axis orientation of the crystal, with a zigzag arrangement of the crystal direction. For squared SnS nanosheets, a similar electrical behavior was found. The resistance along the armchair orientation was found to be 1.6 times higher than that along the zigzag orientation. Recently, our group studied the electrical conductivity of micron-scale colloidal SnS nanosheets.[45] By changing the angle of the four contacts, we successfully detected the *I-V* performances along either the anisotropic directions (zigzag and armchair) or the isotropic directions (**Figure 5e** and **5f**). A large difference in conductivity along the two different anisotropic crystalline directions (1.67:1) was demonstrated, which agreed well with the results reported by Biacchi *et al.*

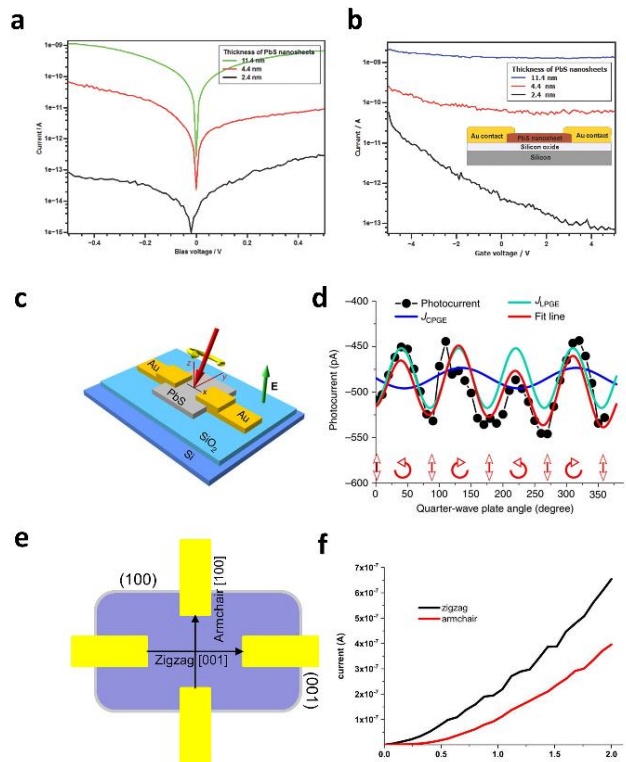

**Figure 5:** (a,b) Nanosheet thickness-dependent I-V characteristics. Reproduced with permission from ref.[59] Copyright 2015 Wiley-VCH. (c) Schematic of a PbS nanosheet-based spintronic device. (d) Generation of circular photo-galvanic effect (CPGE) due to SOC. Reproduced with permission from ref.[4] Copyright 2017 the authors. Published by Springer Nature under a Creative Commons Attribution 4.0 International License http://creativecommons.org/licenses/by/4.0/. (e) Schematic and (f) I-V characteristics of an individual SnS naosheet-based device with electrodes aligning to the anisotropic (zigzag/armchair) direction. Reproduced with permission from ref.[45] Copyright 2019 American Chemical Society.

## 5. CONCLUSIONS AND OUTLOOK

With the rapid development of solution-based synthesis technique in the past decades, a significant progress in the fabrication of colloidal two-dimensional metal chalcogenides had been made. Researchers are now able to precisely tune the geometrical parameters (i.e., crystal shape, thickness, lateral size, shell and crown structure, degree of stacking), stoichiometric ratio and phase composition of two-dimensional metal chalcogenide nanocrystals. Besides the Cd-based nanoplatelets which display their optical signatures typically across the ultraviolet-visible window, nanocrystals with optical response at infrared wavelengths (e.g., Pb-, Hg- and Cu-based chalcogenides) and in-plane anisotropic properties (e.g., SnX, X=S or Se) had also been developed.

By electrical transport characterizations, fascinating properties of individual colloidal nanoplatelet and nanosheet were revealed, such as structure- and spin-dependent optoelectronic response. The discoveries shed light on their tremendous potential of being used as building blocks in electronics and photonics, including surface-defined materials with topological states for spin transport,[49] low-threshold pulsed or continuous wave lasers,[84] quantum sensors,[85] light-emitting diodes and solar cells.

Since colloidal 2D CMCs nanocrystals are solution-processable, it is possible to fabricate flexible electronic devices and lab-on-chip systems by facial steps of spin-coating or inkjet printing in combination with microfabrication. Nevertheless, improving the stability of colloidal nanomaterials against oxidation on an individual crystal scale remains a challenge, which requires further investigations.

## AUTHOR INFORMATION

**Corresponding Author**

* Corresponding author: christian.klinke@uni-rostock.de

**Present Addresses**

†Department of Physics & Warwick Centre for Analytical Science, University of Warwick, Coventry, CV4 7AL, United Kingdom

**Author Contributions**

The manuscript was written, revised, and discussed by all authors, whereas Ziyi Hu contributed the largest part.

**Notes**

The authors declare no competing financial interest.

## Biographical Information

**Ziyi Hu** obtained his master's degree in material chemistry from the Northeast Petroleum University (P.R. China). Now, he is a PhD student at the Department of Physics, University of Warwick. He carries out works in Warwick Center for Ultrafast Spectroscopy and his current research is to investigate the optical properties of low-dimensional metal chalcogenides by ultrafast pump-probe spectroscopy.

**Ryan James Patrick O'Neill** obtained his bachelor's degree in Chemical, Process and Energy Engineering and proceeded to complete a research masters in fuel technology in the Energy Safety Research Institute (ESRI) of Swansea University. From here, he embarked on a PhD within the Chemistry Department of Swansea University, researching biomass valorization through the use of photo-electrocatalysis methods.

**Rostyslav Lesyuk** studied at the Lviv Polytechnic National University where he obtained his master's degree in lasers and optoelectronics and the PhD degree at the Department of Photonics. Then, he was a Post-Doc at the North Dakota State University (ND, USA) to study of Cd-free quantum dots for optoelectronics, supported by the Fulbright Faculty Development Program Scholarship. He proceeded for a further post-doctoral research at Dresden University (Dresden Fellowship program). Later in 2016 he joined the group Prof. C. Klinke at the University of Hamburg to study colloidal 2D systems. Since 2019, Dr. Lesyuk is working as scientific fellow at University of Rostock, Institute of Physics.

**Christian Klinke** studied physics at the University of Würzburg and the University of Karlsruhe (Germany) where he also obtained his diploma degree. He obtained his PhD at the EPFL (Lausanne, Switzerland). Then, he worked as Post-Doc at the IBM TJ Watson Research Center (Yorktown Heights, USA). Thereafter, he started as assistant professor at the University of Hamburg. Since 2017, he is an associate professor at the [Chemistry Department](#) of the [Swansea University](#) and since 2019 full professor at the [Institute of Physics](#) of the [University of Rostock](#). His research concerns the colloidal synthesis of nanomaterials and the optoelectronic characterization of these materials.

## ACKNOWLEDGMENT

This work was supported by a fellowship of the China Scholarship Council (File No.: 201808230269) and an EPSRC 2016 Doctoral Training Partnership award (Ref.: EP/N509553/1).

**Conspectus graphic**

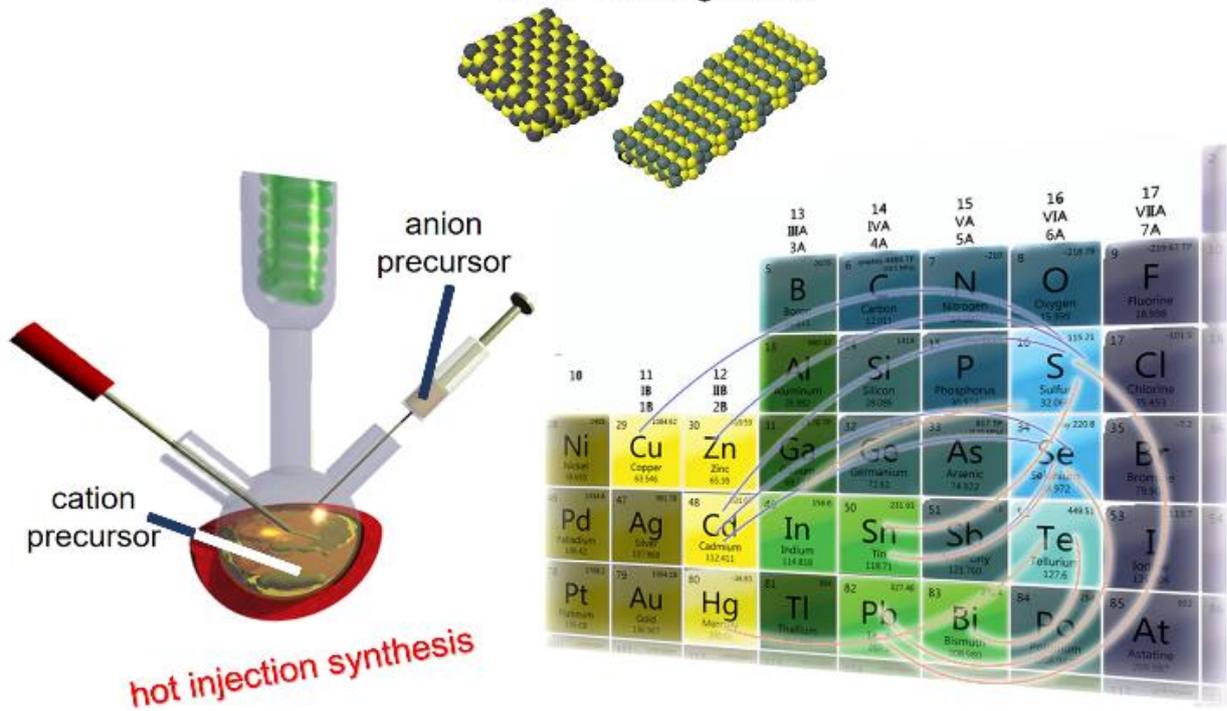